\documentclass[a4paper,fleqn,usenatbib]{mnras}

\usepackage[T1]{fontenc}
\usepackage{ae,aecompl}
\usepackage{graphicx,times}
\usepackage{amsmath,amssymb,bm}
\usepackage{multirow, multicol}
\usepackage{url}
\usepackage{natbib}

\title[The Correspondence between Peaks and Haloes]{The Correspondence between Convergence Peaks from Weak Lensing and Massive Dark Matter Haloes}

\author[C.Wei et al.]{Chengliang Wei,$^{1,2,3}$\thanks{E-mail: chengliangwei@pmo.ac.cn},
Guoliang Li$^{1,2}$, Xi Kang$^{1,2}$, Xiangkun Liu$^{4}$, Zuhui Fan$^{5}$,
\newauthor
Shuo Yuan$^{5}$, Chuzhong Pan$^{5}$ \\
$^{1}$Purple Mountain Observatory, the Partner Group of MPI f\"ur Astronomie, 8 Yuanhua Road, Nanjing 210034, China \\
$^{2}$School of Astronomy and Space Sciences, University of Science and Technology of China, Hefei 230029, China \\
$^{3}$Graduate School, University of the Chinese Academy of Science, 19A Yuquan Road, Beijing 100049, China \\
$^{4}$South-Western Institute for Astronomy Research, Yunnan University, Kunming 650500, China \\
$^{5}$Department of Astronomy, School of Physics, Peking University, Beijing 100871, China
}
\date{Accepted XXX. Received YYY; in original form ZZZ}
\pubyear{2018}

\begin{document}
\label{firstpage}
\pagerange{\pageref{firstpage}--\pageref{lastpage}}
\maketitle

\begin{abstract}
     The convergence peaks, constructed from galaxy shape measurement in weak lensing, 
     is a powerful probe of cosmology as the peaks can be connected with the underlined 
     dark matter haloes. However the capability of convergence peak statistic is affected 
     by the noise in galaxy shape measurement, signal to noise ratio as well as the 
     contribution from the projected mass distribution from the large-scale structures 
     along the line of sight (LOS). In this paper we use the ray-tracing simulation on a 
     curved sky to investigate the correspondence between the convergence peak and the 
     dark matter haloes at the LOS. We find that, in case of no noise and for source galaxies 
     at $z_{\rm s}=1$, more than $65\%$ peaks with $\text{SNR} \geq 3$ (signal to noise ratio) are related to more 
     than one massive haloes with mass larger than $10^{13} {\rm M}_{\sun}$. Those massive 
     haloes contribute $87.2\%$ to high peaks ($\text{SNR} \geq 5$) with the remaining contributions 
     are from the large-scale structures. On the other hand, the peaks distribution is skewed 
     by the noise in galaxy shape measurement, especially for lower SNR peaks. In the noisy 
     field where the shape noise is modelled as a Gaussian distribution, about $60\%$ high peaks 
     ($\text{SNR} \geq 5$) are true peaks and the fraction decreases to  $20\%$ for lower peaks ($ 3 \leq \text{SNR} < 5$). 
     Furthermore, we find that high peaks ($\text{SNR} \geq 5$) are dominated by very massive haloes 
     larger than $10^{14} {\rm M}_{\sun}$.

\end{abstract}


%%%%%%%%%%%%%%%%%%%%%%%%%%%%%%%%%%%%%%%%%%%%%%%%%%
%%%%%%%%%%%%%%%%% BODY OF PAPER %%%%%%%%%%%%%%%%%%

%%%%%%%%%%%%%%%%% Introduction
\section{INTRODUCTION}
\label{sect:intro}
In the context of weak lensing, light ray of background source is deflected by
the foreground mass and usually it leads to a weak distortion of the image of
the background source \citep[][]{1992grle.book.....S, 2001PhR...340..291B,
2006glsw.conf.....M, 2017SchpJ..1232440B}. The distortion of galaxy image is sensitive to the structure
growth of the universe, providing us with a promising method to directly probe
the distribution of matter density and the geometry of the universe \citep[][]{
1999ARA&A..37..127M, 2001A&A...374..757V, 2003astro.ph..9482K, 2008JCAP...04..013A,
2008A&A...479....9F, 2015RPPh...78h6901K, 2016MNRAS.463.3326F}. Weak lensing
statistic has been one of the main scientific goal of ongoing and upcoming galaxy
surveys, such as the Kilo-Degree Survey \citep[KiDS, ][]{2015A&A...582A..62D}, Dark
Energy Survey \citep[DES, ][]{2016MNRAS.460.1270D}, Euclid Mission
\citep[][]{2011arXiv1110.3193L} and the Large Synoptic Survey Telescope
\citep[LSST, ][]{2009arXiv0912.0201L}.

To extract cosmological information, the shear correlation functions have been widely
used (e.g., \citealt{2013MNRAS.432.2433H, 2016ApJ...824...77J, 2017MNRAS.465.1454H};
see \citealt[][]{2015RPPh...78h6901K} for a review). The shear correlation function 
statistics is straightforward but it is also clear that it  
 involves the galaxy intrinsic alignment which is a major systematic
contribution and has to be carefully calibrated \citep[e.g., ][]{2018ApJ...853...25W}.
As a complementary statistics to the shear correlation, the convergence peak, a direct
probe of massive dark matter haloes \citep[e.g., ][]{2004MNRAS.350..893H, 2005A&A...442..851M},
is another useful approach to probe cosmology by measuring the number of lensing peaks
\citep[e.g., ][]{2010MNRAS.402.1049D, 2010PhRvD..81d3519K, 2011MNRAS.416.2527M,
2011PhRvD..84d3529Y, 2013MNRAS.430.2896C, 2013MNRAS.432.1338M, 2015A&A...576A..24L,
2015MNRAS.450.2888L,2016PhRvL.117e1101L, 2015MNRAS.453.3043S, 2017MNRAS.465.1974S,
2018arXiv180101886G}. It has been shown that the convergence peaks can be used to break
the degeneracy between the different parameters (e.g., $\Omega_{\rm m}$ and $\sigma_{8}$) 
\citep{2018MNRAS.474.1116S}. Using the recent survey data from KiDS-450 \citep{2017MNRAS.465.1454H}, 
\citet{2018MNRAS.474..712M} have confirmed that combining with the measurements of
convergence peaks can improve $\sim 20\%$ in the uncertainty of $S_{8}$ compared to the
non-tomographic measurements of the shear two-point correlation functions alone.

Lensing peaks are defined as the local maxima in the reconstructed convergence
field. It has been shown that haloes are important contributors to the convergence peaks. High
peak, with high signal-to-noise ratio, is usually caused by one single massive halo along
the line of sight \citep[LOS, e.g., ][]{ 2004MNRAS.350..893H, 2014MNRAS.442.2534S, 2015MNRAS.453.3043S},
while low peak with lower-significance could be contributed by a few low mass haloes along the LOS
\citep[e.g.,][]{2011PhRvD..84d3529Y, 2016PhRvD..94d3533L}. It is also found that peaks in
the convergence map have other origins, such as the projection of LSS, shape noise of
galaxies, or mask effects in the observation \citep{2000MNRAS.313..524V, 2002ApJ...575..640W, 2005A&A...442..851M,
2010ApJ...719.1408F, 2011PhRvD..84d3529Y, 2014ApJ...784...31L, 2018ApJ...857..112Y}.
Our understanding is not yet complete, especially for the medium and low peaks \citep[e.g., ][]{2016PhRvD..94h3506Z}.

Observationally, weak lensing convergence peak has been used to detect galalxy cluster in
recent galaxy imaging survey \citep{2007ApJ...669..714M, 2015ApJ...807...22M, 2012ApJ...748...56S}.
In \citet{2015ApJ...807...22M}, they checked the one-to-one correspondence between convergence peaks
and clusters, and found that the number of peaks identified in the Hyper Suprime-Cam's $2.3\ {\rm deg}^2$
field is considerably larger than the average number expected from $\Lambda$CDM cosmology. They
explained that this difference is due to the large sample variance in the small field of galaxy survey.
Moreover, in \citet{2012MNRAS.425.2287H} they studied the scatter and bias in weak lensing selected
clusters, and concluded that the SNR of a convergence peak is not a good indicator of halo virial mass but
is more tightly related to the inner mass such as $M_{1000}$.
In this paper, we use the mock data of \citet{2018ApJ...853...25W} to investigate the correspondence
between convergence peaks and dark matter haloes, and the effects of contamination by galaxy shape noise.

To date, full-sky weak-lensing maps have been produced in several works
\citep{2008MNRAS.391..435F, 2015MNRAS.447.1319F, 2009A&A...497..335T,  2013MNRAS.435..115B, 2015MNRAS.453.3043S}.
These simulations covered a sufficient large volume to compute a full-sky convergence (and shear) maps,
and measure lensing powers from both the linear and nonlinear regime. In this paper, we follow
\citet{2018ApJ...853...25W} to simulate weak lensing maps on the full-sky, involving a wide range for
both mass and length scales, which is non-trivial for the future large-area sky surveys \citep{2017MNRAS.472.2126K, 2017JCAP...05..014L}.
Using the simulated lensing maps, we explore the correspondence between convergence peaks and massive haloes,
and also examine the contribution of LSS to a given lensing peak with different peak heights.

The rest of this paper is organized as follows. In Sec. 2, we first introduce the necessary
background of weak lensing, and describe the basics of convergence peak. Sec. 3 describes
our simulation and the spherical ray-tracing technique. In Sec. 4, we present the results
of peak statistics in our measurements. Conclusions and discussions are given in Sec. 5.

%%%%%%%%%%%%%%%%% Background
\section{WEAK GRAVITATIONAL LENSING}
\label{sect:Obs}
In this  section, we briefly introduce the theoretical background
of  weak  gravitational  lensing, and also describe some basics on
convergence peak analysis.

%%%%%%%%%%%%%%%%%%%%%%%%%%%%%%%%%%%%%%%%
\subsection{Lensing Basics}
In the weak lensing context, light ray, located at $\bm{\beta}$ on
source plane, is deflected by the intervening mass field along the
LOS as \citep{2001PhR...340..291B, 2006glsw.conf.....M, 2015RPPh...78h6901K}
\begin{equation}
  \beta_{i}(\bm{\theta}, \chi) = \theta_{i} -
  \frac{2}{c^2}\int^{\chi}_{0}{\rm d\chi'}
  \frac{r(\chi-\chi')}{r(\chi)r(\chi')}
  \Phi_{,i}\left( r(\chi')\bm{\theta}, \chi' \right),
\end{equation}
where $c$ is the speed of light, $\bm{\theta}$ represents the
observed position of the given light. $\Phi_{,i}$ is  spatial
derivative of the gravitational potential on the path of light
rays.  $r(\chi)$  is  the  comoving angular diameter distance,
which is equal to the comoving  distance  $\chi$  in  a  flat
cosmology. In the weak lensing regime, lensing distortion of
light beams can be quantified by the Jacobian matrix to linear
order,
\begin{equation}
\begin{aligned}
  \mathcal{A}_{ij}(\bm{\theta}, \chi)
  & =\frac{\partial \beta_{i} (\bm{\theta}, \chi)}{\partial \theta_{j}} \\
  & =\delta_{ij}-\frac{2}{c^2}\int_{0}^{\chi} {\rm d}\chi'
     \frac{r(\chi-\chi')r(\chi')}{r(\chi)}\Phi_{,ij}(r(\chi')\bm{\theta},\chi').
\end{aligned}
\end{equation}
Here,  $\delta_{ij}$  is  Kronecker  delta.  Typically,  this
distortion  matrix  $\mathcal{A}(\bm{\theta}, \chi)$  can  be
decomposed into several components as
\begin{equation}
  \mathcal{A}(\bm{\theta}, \chi)=\begin{pmatrix}1-\kappa-\gamma_1 & -\gamma_2 \\
                                 -\gamma_2 & 1-\kappa+\gamma_1 \end{pmatrix},
\end{equation}
where $\kappa$ is the convergence field, describing an isotropic
stretching       of       the   image.   While the complex shear,
$\bm{\gamma}=\gamma_{1}+{\rm i} \gamma_{2}$, represents the change
of shape between the image and  the  source. The gravitational
potential $\Phi$ can be related to the density contrast $\delta$
by the Poisson equation. Hence, lensing convergence, which is the
projected matter density field, can be given as
\begin{equation}
  \kappa(\bm{\theta}, \chi) = \frac{3H_{0}^{2}\Omega_{\rm m}}{2c^2}
  \int_{0}^{\chi}{\rm d}\chi'\frac{r(\chi-\chi')r(\chi')}{r(\chi)}
  \frac{\delta(r(\chi')\bm{\theta},\chi')}{a(\chi')},
  \label{equ:kappa}
\end{equation}
where $H_{0}$  is the present Hubble  constant, $\Omega_{\rm m}$
is  the  matter  density  in  units  of  the  critical density, and
$a(\chi')$ is the scale factor at $\chi'$.

%%%%%%%%%%%%%%%%%%%%%%%%%%%%%%%%%%%%%%%%
\subsection{Convergence Peak}
To use the peak convergence as a cosmological probe, one needs to
reconstruct the convergence from the shear field \citep[e.g.,][]{
1999A&A...348...38L, 2017ApJ...839...25S}. However, due to the shape
noise in the measurements, the reconstructed convergence field will
be contaminated by an additive noise field as \citep{2004MNRAS.350..893H,
2010ApJ...719.1408F}
\begin{equation}
  \kappa_{\rm n}(\bm{\theta}) = \kappa({\bm{\theta}}) + n(\bm{\theta}).
\end{equation}
Therefore, to suppress the measurement noise, the convergence field is
usually smoothed as
\begin{equation}
  \mathcal{K}_{\mathcal{N}}(\bm{\theta})
  = \mathcal{K}(\bm{\theta}) + \mathcal{N}(\bm{\theta})
  = \int {\rm d}^{2} \bm{\theta}'
    \kappa_{\rm n}(\bm{\theta} - \bm{\theta}') U(\bm{\theta}').
\end{equation}
Here the window function $U(\theta)$ is usually taken as a Gaussian
smoothing filter \citep{2004MNRAS.350..893H, 2015MNRAS.453.3043S}
\begin{equation}
  U(\theta)= \frac{1}{\pi \theta_{\rm G}^{2}}
             \exp \left( -\frac{\theta^{2}}{\theta_{\rm G}^{2}} \right),
\end{equation}
where smoothing scale $\theta_{\rm G}$ is set to be $2\ {\rm arcmin}$ in
the following analyses, which is a typical scale for weak lensing cluster
detection with sources at $z_{\rm s} = 1.0$.

As  proposed  by  \citet{2000MNRAS.313..524V},  galaxy shape noise
$\mathcal{N}(\bm{\theta})$ in the lensing measurement can be given
as a Gaussian random field with the variance as \citep[e.g., ][]{2000ApJ...530L...1J}
\begin{equation}
  \sigma_{\mathcal{N}}^{2} = \frac{\sigma_{\epsilon}^{2}}{4\pi n_{\rm g} \theta_{\rm G}^{2} },
\end{equation}
where $n_{\rm g}$ and $\sigma_{\epsilon}^{2}$ denote the number density
and the intrinsic ellipticity dispersion of source galaxies, respectively.
In this work, shape noise is added to the simulated lensing maps with
$\sigma_{\epsilon} = 0.4$ and $n_{\rm g} = 10\ {\rm arcmin}^{-2}$ for a
CFHTLenS-like survey, which means $\sigma_{\mathcal{N}} \approx 0.018$ for
our choice. With the purpose of investigating lensing peaks in the convergence
maps, we can extract convergence peaks by defining the local maxima or minima
from the field of the signal-to-noise ratio (SNR),
\begin{equation}
  \nu(\bm{\theta}) = \frac{\mathcal{K}_{\mathcal{N}} (\bm{\theta})}{\sigma_{\mathcal{N}}}.
\end{equation}
Hereafter we adopt the same $\sigma_{\mathcal{N}}$ to calculate SNR for both
noise convergence field and noise-free field. In this paper, we use two thresholds
$\kappa = 3\sigma (0.054)$ and $\kappa = 5\sigma (0.089)$ for our peak statistics.

In general, the convergence peaks with high SNR are caused by massive cluster of
galaxies \citep[e.g.,][]{2004MNRAS.350..893H, 2007ApJ...669..714M, 2015ApJ...807...22M,
2012ApJ...748...56S, 2014MNRAS.442.2534S}.  For this reason, one can relate local
high  convergence  peaks with associated massive haloes along the LOS. $N$-body simulations have shown
that  dark  matter  haloes  are well described by the universal NFW density profile
\citep{1996ApJ...462..563N, 1997ApJ...490..493N},
\begin{equation}
  \rho_{\rm NFW}(r) = \frac{\rho_{\rm s}}{(r/r_{\rm s})(1+r/r_{\rm s})^{2}},
\end{equation}
Here the scale radius of a given halo is $r_{\rm s} = r_{\rm vir}/\mathcal{C}$,
where  $\mathcal{C}$   is   known   as  the dimensionless concentration parameter
of a halo, which can be related to the halo mass $M_{\rm vir}$ as
\begin{equation}
  \mathcal{C}(M_{\rm vir}, z) = \alpha \left( \frac{M_{\rm vir}}{M_{\rm pivot}} \right)^{\beta} (1+z)^{\gamma}
\end{equation}
In  		this	 	paper,	 	we 	use the fitting parameters
$(\alpha,\ \beta,\ \gamma,\ M_{\rm pivot}) = (5.72,\ -0.081,\ -0.71,\ 10^{14}h^{-1}{\rm M}_{\sun})$
given in \citet{2008MNRAS.390L..64D}. The scale density $\rho_{\rm s}$ is defined as
$\rho_{\rm s} = \delta_{\rm c} \rho_{\rm c}$, where $\rho_{\rm c}$  is  the  critical
density  of  the universe, and the characteristic overdensity $\delta_{\rm c}$ of the
halo is
\begin{equation}
  \delta_{\rm c} = \frac{\delta_{\rm vir}}{3}
                   \frac{\mathcal{C}^{3}}{\ln(1+\mathcal{C})-\mathcal{C}/(1+\mathcal{C})},
\end{equation}
where $\delta_{\rm vir}$ is the threshold overdensity for
spherical  collapse.  Therefore,  $\rho_{\rm s}$  can  be
related to $M_{\rm vir}$ as
\begin{equation}
\begin{aligned}
M_{\rm vir} & = \delta_{\rm vir} \frac{4\pi}{3} r_{\rm vir}^{3} \rho_{\rm c} \\
            & = 4\pi \rho_{\rm s} \left( \frac{r_{\rm vir} }{\mathcal{C}_{\rm vir}} \right)^{3}
                \left[ \ln(1+\mathcal{C})-\frac{\mathcal{C}}{1+\mathcal{C}} \right].
\end{aligned}
\end{equation}

With the thin lens approximation, the surface mass density is obtained
by integrating the density profile along the LOS. Then the dimensionless
convergence is given as \citep{2003MNRAS.340..580T}
\begin{equation} \label{equ:kNFW}
  \kappa_{\rm NFW}(x) = \frac{2\rho_{\rm s} r_{\rm s}}{\Sigma_{\rm c}} \mathcal{F}(x),
\end{equation}
with
\begin{equation}
  \Sigma_{\rm c} = \frac{c^{2}}{4\pi G} \frac{D_{\rm s}}{D_{\rm l}D_{\rm ls}},
\end{equation}
where we defined the dimensionless radius $x = r/r_{\rm s}$.
$D_{\rm l}$,  $D_{\rm s}$  and  $D_{\rm ls}$  are the
angular dimeter distance to the lens, to the source and between the lens
and the source, respectively. The factor $\mathcal{F}(x)$ is
\begin{equation}
  \mathcal{F}(x) = \left\{
  \begin{array}{lcl}
    \frac{\sqrt{\mathcal{C}^2-x^2}}{(x^2-1)(\mathcal{C}+1)}
      +\frac{1}{(1-x^2)^{3/2}}{{\rm arccosh}} \left[\frac{x^2+\mathcal{C}}{x(\mathcal{C}+1)}\right]
      \\  \hfill x < 1, \\

    \frac{(\mathcal{C}+2) \sqrt{\mathcal{C}^{2}-1} }{3(\mathcal{C}+1)^{2}}  \hfill  x = 1, \\

    \frac{\sqrt{\mathcal{C}^2-x^2}}{(x^2-1)(\mathcal{C}+1)}
      -\frac{1}{(x^2-1)^{3/2}}{\rm arccos}\left[\frac{x^2+\mathcal{C}}{x(\mathcal{C}+1)}\right]
      \\ \hfill  1 < x \leq \mathcal{C}, \\

    0  \hfill x > \mathcal{C}.
  \end{array}
  \right.
\end{equation}
Here the density profile is truncated at the virial radius of the NFW halo.

%%%%%%%%%%%%%%%%% Simulation
\section{NUMERICAL SIMULATION}
In this  section,  we  introduce our $N$-body simulation and the
spherical  ray-tracing technique used in our lensing simulation.

%%%%%%%%%%%%%%%%%%%%%%%%%%%%%%%%%%%%%%%%
\subsection{$N$-body Simulation}
In \citet{2018ApJ...853...25W} the distribution of cosmological
matter density is simulated by $3072^{3}$ dark matter particles
in a cubic box of $L_{\rm box}=500h^{-1}{\rm Mpc}$ on each side
with {\tt L-GADGET}, a memory-optimized version of {\tt GADGET2}
\citep{2005MNRAS.364.1105S}. This simulation, referred as 'L500',
is run as a part of the ELUCID project \citep{2014ApJ...794...94W,
2016ApJ...831..164W, 2016RAA....16..130L, 2017ApJ...841...55T} with
WMAP9 cosmology \citep{ 2013ApJS..208...19H}: $\Omega_{\rm m} = 0.282$,
$\Omega_{\Lambda} = 0.718$, $\Omega_{\rm b} = 0.046$, $h = 0.697$,
$\sigma_{8} = 0.817$, and $n_{\rm s}=0.965$. Each particle in the
simulation has a mass of $3.375 \times 10^{8} h^{-1} {\rm M}_{\sun}$.
Dark matter haloes in the simulation are identified by the
friends-of-friends (FOF) algorithm with $b = 0.2$, which  is the
linking parameter in units of the mean particle separation.

To trace back the light beams in our mock cosmology, we need
to employ our $N$-body simulation to build a past light-cone.
In \citet{2018ApJ...853...25W}, they first divide the large
simulation box into sets of small cubic boxes with
$100 h^{-1} {\rm Mpc}$ on each side. An observer is randomly
located at an off-center position at $z_{0}$. In terms of the
redshifts, these cell boxes are appropriately piled together
so as to  cover  the past light-cone from $z_{0}$ to $z_{\rm max}$.
Then a full-sky light-cone can be constructed to redshift
$z_{\rm max}$ in this manner from the L500 simulation.

%%%%%%%%%%%%%%%%%%%%%%%%%%%%%%%%%%%%%%%%
\subsection{Spherical Ray-tracing Simulation}
To produce lensing map from simulation, we follow the multi-plane
algorithm on the spherical geometry, developed by \citet{2008ApJ...682....1D},
\citet{2008MNRAS.391..435F}, \citet{2009A&A...497..335T},
\citet{2013MNRAS.435..115B}. In this section we will briefly summarize
the main procedures of our lensing simulation. More details can be found
in \citet{2018ApJ...853...25W}.

To perform ray-tracing simulation on the curved sky,
the light-cone is then decomposed into a set of spherical shells  of
a given width, $50\ h^{-1}{\rm Mpc}$, around the observer, and the
dark matter particles are projected to the corresponding shells, pixelized
by the HEALPix \footnote{\url{http://healpix.jpl.nasa.gov}}
tessellation \citep{2005ApJ...622..759G, 2007MNRAS.381..865C}. In this work,
the HEALPix resolution parameter is set to $N_{\rm side} = 8192$, which gives
an angular resolution of ${\rm d}\theta \sim 0.43\ {\rm arcmin}$ \footnote{The
pixel size of HEALPix cells can be calculated by
${\rm d}\theta = \left( 4\pi/12N_{\rm side}^{2}\right)^{1/2}$ for a given
$N_{\rm side}$.}. Then the projected surface mass densities are calculated for
each shell using the SPH smoothing algorithm \citep{2005ApJ...635..795L,
2010ARA&A..48..391S}. We use the nearest 64 particles to define the kernel size,
but keep this smoothing length larger than two HEALPix cells in the high-density
regions. Finally we can obtain the surface matter overdensity $\Sigma^{(n)}$
on the $n$-th shell by
\begin{equation}
  \Sigma^{(n)}(\bm{\theta}^{(n)}) = \int^{\chi_{n+1/2}}_{\chi_{n-1/2}}
                                    {\rm d}\chi' \delta(r(\chi')\bm{\theta}^{(n)}, \chi').
\end{equation}
where $\delta = \rho/\bar{\rho} - 1$ is the mass overdensity. The convergence field
is then given by
\begin{equation}
  \kappa^{(n)}(\bm{\theta}^{(n)})  = W^{(n)}\Sigma^{(n)}(\bm{\theta}^{(n)}),
\end{equation}
where lensing kernel function $W^{(n)}$ is defined as
\begin{equation}
  W^{(n)} = \frac{3}{2}\Omega_{\rm m} \left( \frac{H_{0}}{c} \right)^{2}
            \frac{r(\chi_{n})}{a(\chi_{n})}.
\end{equation}
Using the Poisson equation, one can obtain the lensing potential of
the $n$-th shell $\phi_{\ell m}^{(n)}$ from the mass  density  shell
in the spherical harmonic space
\begin{equation}
  \phi_{\ell m}^{(n)} = -\frac{2}{\ell (\ell+1)} \kappa_{\ell m}^{(n)}.
  %\nabla^2 \phi^{(n)} = 2\kappa^{(n)}
\end{equation}
where $\kappa_{\ell m}^{(n)}$ are the spherical harmonic coefficients of
the convergence $\kappa^{(n)}$. The terms of $\kappa_{00}^{(n)}$ correspond
to the mean of convergence map. They have no effects on the deflection angle
and distortion. The deflection field $\bm{\alpha}^{(n)}$ can be derived
from the gravitational lensing potential as,
\begin{equation}
  %\alpha_{\ell m}^{(n)} = -\sqrt{\ell(\ell+1)} \phi_{\ell m}^{(n)}
  \bm{\alpha}^{(n)}=\nabla \phi^{(n)}.
\end{equation}
Therefore, light rays are initialized at the center of each HEALPix cell in the
first shell, and then will be propagated to the next lens plane following
\citep[e.g., ][]{2009A&A...497..335T, 2013MNRAS.435..115B}
\begin{equation}
  \bm{x}^{(n+1)} = \mathcal{R}(\bm{n}^{(n)}\times\bm{\alpha}^{(n)},
                   \lVert\bm{\alpha}^{(n)}\rVert)  \bm{x}^{(n)},
\end{equation}
where $\bm{n}$ denotes the radial direction of light rays on the $n$-th shell,
and the rotation matrix $\mathcal{R}$ makes a rotation with an angle
$\lVert\bm{\alpha}\rVert$ to the light rays. The lensing distortion matrix
$\mathcal{A}^{(n+1)}$ can be calculated at the position of light rays by
\begin{equation}
\begin{aligned}
  \mathcal{A}^{(n+1)}_{ij}
  & = \left(1-\frac{D^{n}_{0}}{D^{n+1}_{0}}\frac{D^{n+1}_{n-1}}{D^{n}_{n-1}} \right) \mathcal{A}^{(n-1)}_{ij}
    + \frac{D^{n}_{0}}{D^{n+1}_{0}}\frac{D^{n+1}_{n-1}}{D^{n}_{n-1}} \mathcal{A}^{(n)}_{ij} \\
  & - \frac{D^{n+1}_{n}}{D^{n+1}_{0}} \mathcal{U}^{(n)}_{ik}\mathcal{A}^{(n)}_{ij},
\end{aligned}
\end{equation}
where $D^{n+1}_{n}$ is the comoving angular diameter distance,
$D^{n+1}_{n} \equiv r(\chi_{n+1}-\chi_{n})$. $\mathcal{U}^{(n)}_{ij}$
denotes the deformation matrix in $n$-th lens plane, which can be
calculated as $\mathcal{U}^{(n)}_{ij} = \phi^{(n)}_{,ij}$.

\begin{figure}
  \centering
  \includegraphics[width=0.49\textwidth, angle=0]{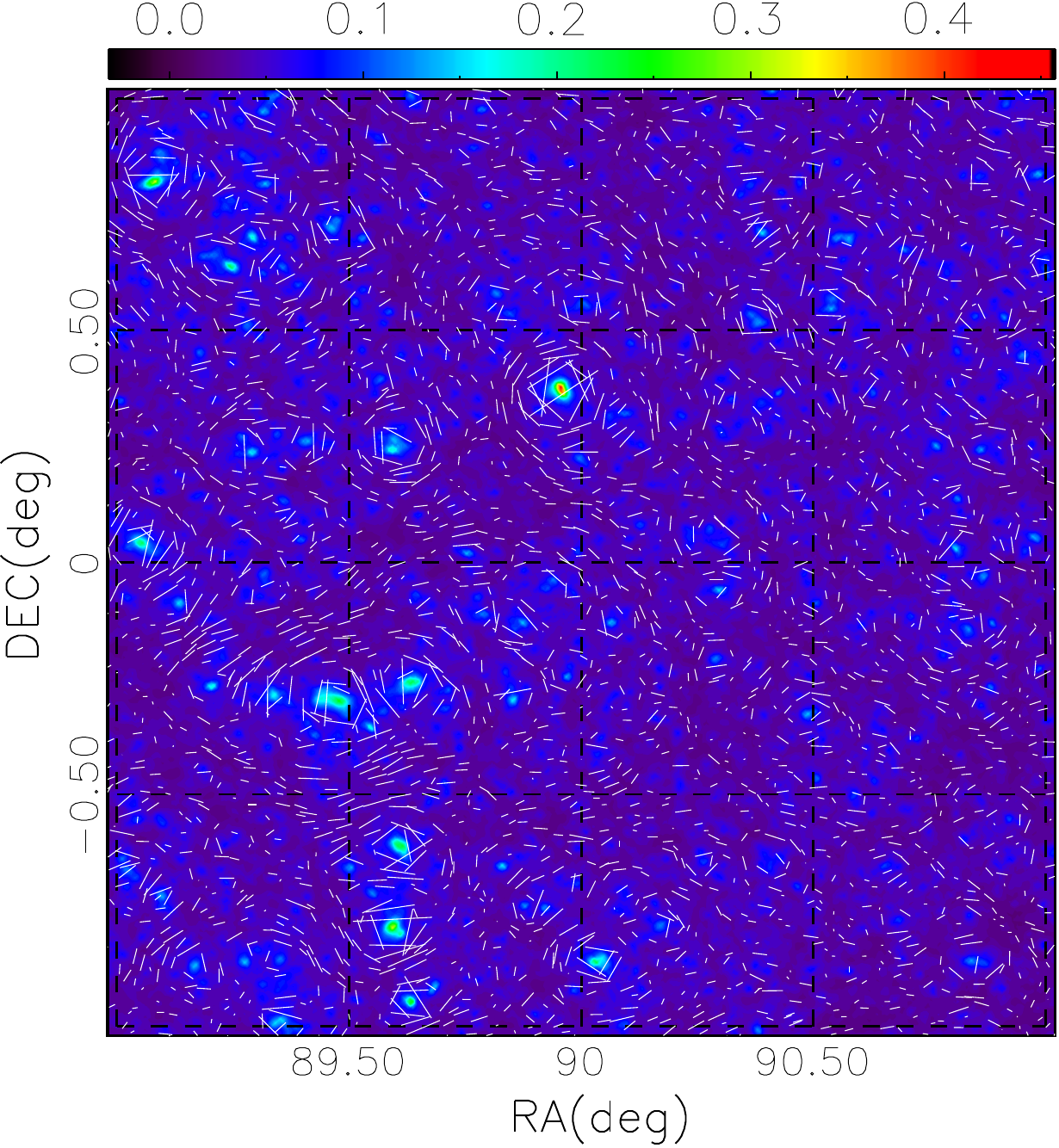}
  \caption{The convergence map in a $2^{\circ} \times 2^{\circ}$ field,
  extracted from one realization of full-sky lensing simulation, for
  sources at $z_{\rm s} = 1.0$. The lensing shear is marked by the short
  sticks.}
  \label{Fig1}
\end{figure}

For more accurate computations in the lensing simulation, we perform the spherical
harmonic analysis with an iterative algorithm as performed in HEALPix subroutine
{\tt map2alm\_iterative}, to control the residual in the solution of the lensing potential.
This operation can remarkably increase the numerical accuracy at small scales \citep{2018ApJ...853...25W}.
Following the spherical ray-tracing simulation, we are able to evaluate the
distortion matrix $\mathcal{A}$ on each lensing shell and construct the lensing
map at a given redshift. As an illustration, Fig.~\ref{Fig1}
displays  one  realization  of  our simulated convergence map in a field of
$2^{\circ} \times 2^{\circ}$ for sources at $z_{\rm s} = 1.0$. In the
figure, the short sticks are overlaped to indicate the lensing shear in the
field. As expected, galaxies will be stretched tangentially around convergence peaks
by the gravitational lensing.

%%%%%%%%%%%%%%%%%%%%%%%%%%%%%%%%%%%%%%%%
\section{CONVERGENCE PEAKS}
\begin{figure}
  \centering
  \includegraphics[width=0.475\textwidth, angle=0]{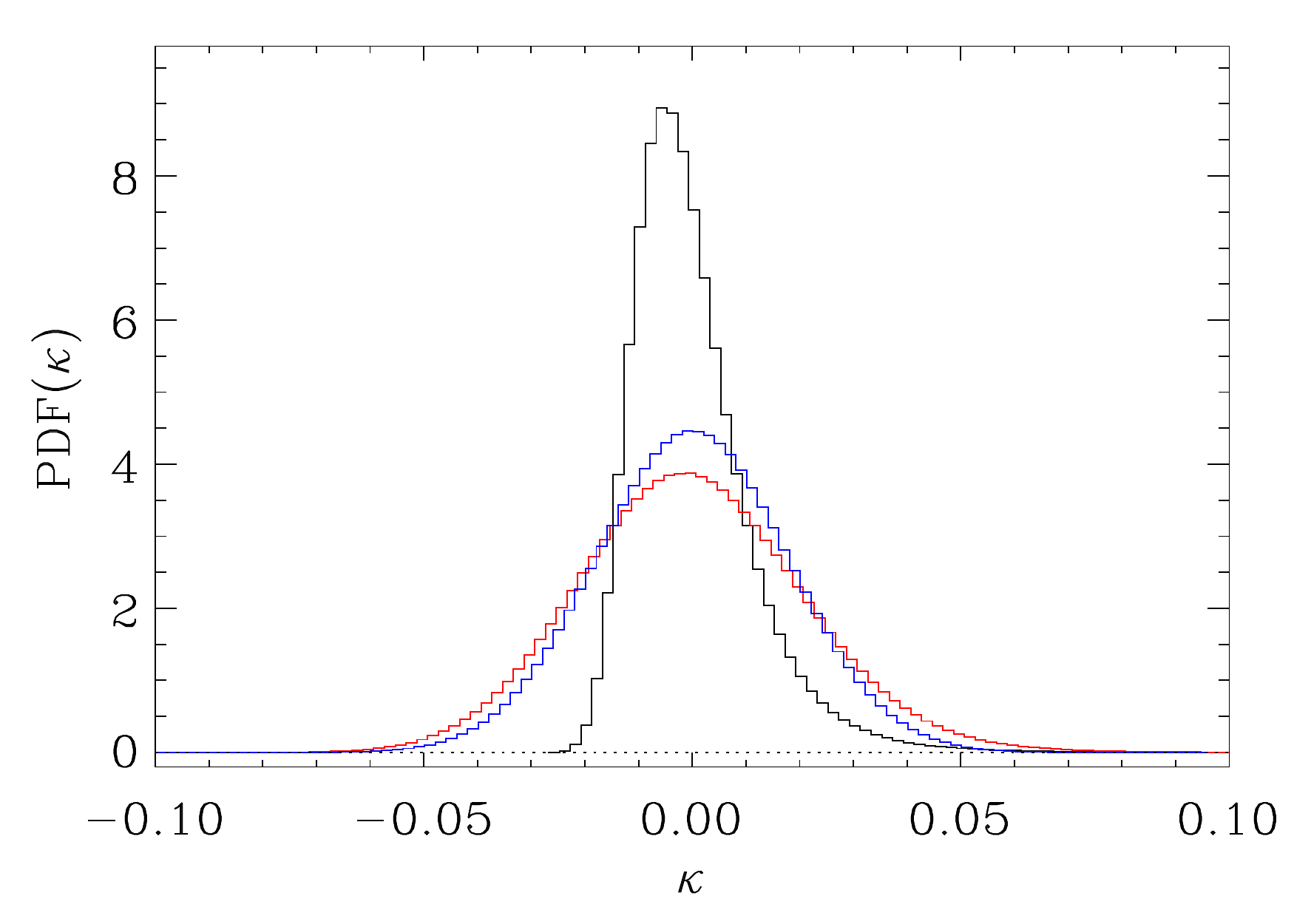}
  \caption{The probability distribution functions of the convergence in noise-free
  fields (black) and noisy fields (red) for $z_{\rm s} = 1$. Blue solid line shows the PDF
  of pure noise with a Gaussian random field.}
  \label{Fig2}
\end{figure}

In this section, we study the statistical properties of lensing peaks based on the simulated
convergence maps within a total coverage of $\sim 2000$ sq. degree, consisting of ten
$\sim 14^{\circ} \times 14^{\circ}$ fields on the sky. The size of each patch is comparable
with CFHTLenS. 
%Note that the size of our simulation box is $500 h^{-1}$Mpc, which is relatively
%small compared with the comoving distance to redshift $z_{\rm s} = 1.0$ and periodic effects
%will show up. To avoid this periodic effects along the lines of sight, these ten sky patches
%are randomly extracted from the full-sky lensing simulation with pointing apart from the
%direction of axes and diagonals of the cartesian coordinates. 
Fig.~\ref{Fig2} shows the
probability distribution functions (PDF) of convergence in the noise-free (black line) and
noisy (red line)  field  for  sources at $z_{\rm s} = 1.0$. As shown in previous work
\citep[e.g.,][]{2000ApJ...530L...1J}, due  to  the  collapsed  massive  haloes, a positive
tail can be found in the noise-free field. There is a cutoff at the negative convergence for
the underdense regions. For the noisy convergence field, it is significantly modulated by the
Gaussian noise. We will use both noise-free and noisy convergence fields for our peak analyses.

\begin{figure}
  \centering
  \includegraphics[width=0.475\textwidth, angle=0]{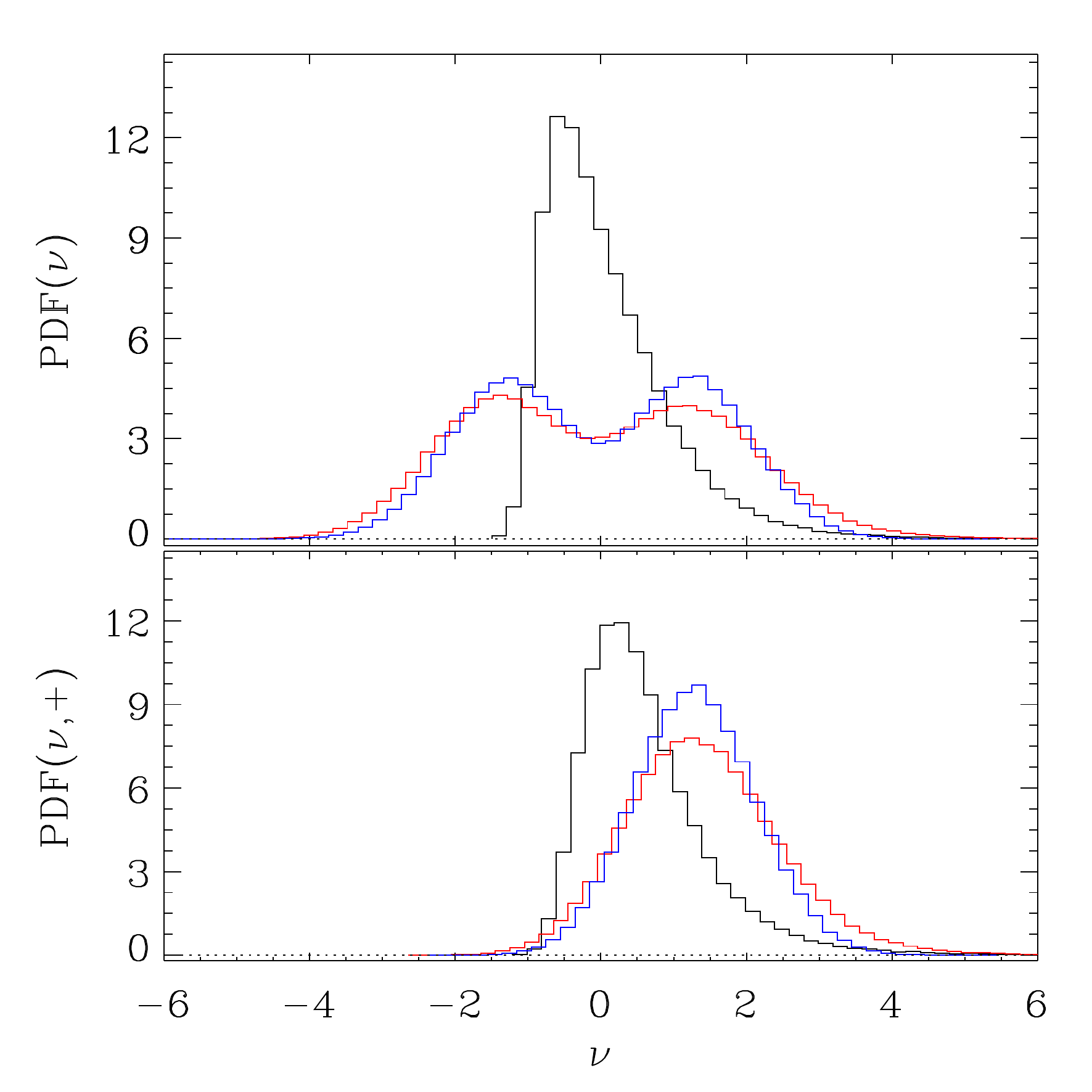}
  \caption{The upper panel shows the SNR distribution of the convergence
  peaks in noise-free fields (black) and noisy fields (red). Blue solid
  line shows the PDF of peaks in the pure noise field. In the lower panel,
  only peaks with the local maxima  in the fields are shown.}
  \label{Fig3}
\end{figure}

%%%%%%%%%%%%%%%%%%%%%%%%%%%%%%%%%%%%%%%%
\subsection{Peak Counts}
As in previous work \citep[e.g., ][]{2015A&A...576A..24L}, the convergence
peaks are defined as those grids which have the local maxima or minima,
meaning their SNR are higher/lower than those adjacent grids. In Fig.~\ref{Fig3},
we show the SNR distribution of the convergence peaks. The top panel shows
peaks with both local maxima and minima and the bottom panel shows those
peaks with local maximum SNR in the convergence field. In the noise-free
case, as shown by the black lines in Fig.~\ref{Fig3}, one can see a tail
at the high-$\nu$ end. It mostly arises from massive haloes, and these
high peaks can be used to study the properties of non-linear gravitational
clustering in the cosmology \citep[e.g.,][]{2000ApJ...530L...1J}. In contrast,
for the noisy convergence maps, red solid line in the upper panel shows that
the asymmetric double-peaks arise due to the noise modulation, and there are
clear excesses on the high peaks at both positive and negative. This implies
that a number of false peaks are produced by the noise. Indeed, providing the
noise field is Gaussian \citep{2000MNRAS.313..524V}, the rate of false peaks,
caused by noise, can be derived from Gaussian random field theory
\citep[e.g.,][]{2010ApJ...719.1408F, 2010A&A...519A..23M}. In the bottom panel
we show the distribution of convergence peaks which are the local maxima. For
the Gaussian noise field, the distribution of local maximum peaks is also
Gaussian as shown by the blue solid line. It is interesting to find some local
maximum peaks with negative height. As discussed in \citet{2004MNRAS.350..893H},
this could happened when a halo resides in an underdense region, such as void
in the space.

\begin{figure*}
  \centering
  \includegraphics[width=0.95\textwidth, angle=0]{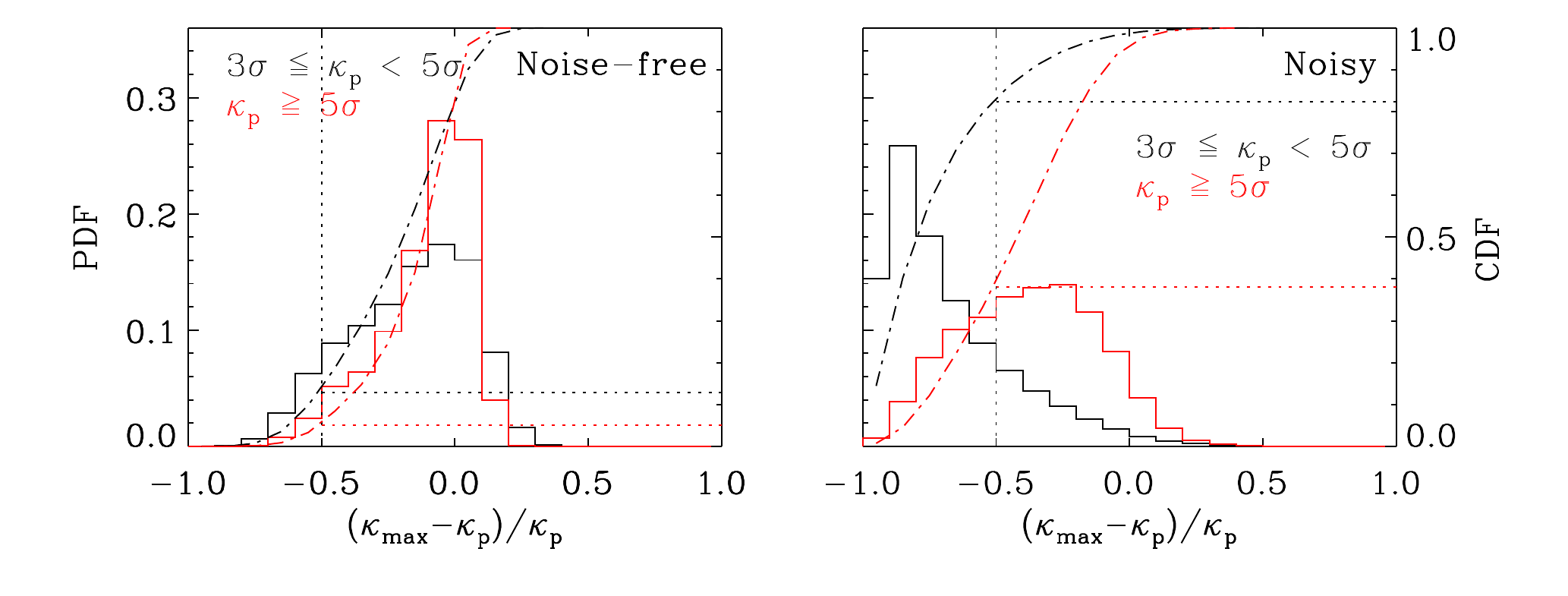}
  \caption{The distribution of the main lens plane contribution for different
  peak heights in noise-free (left) and noisy (right) fields. Here
  $\kappa_{\rm max}$ denotes the maximum contributor of the lens plane to a
  given convergence peak $\kappa_{\rm p}$. In addition, dot-dashed lines show
  the CDF of the fractional difference for the lensing peaks.}
  \label{Fig4}
\end{figure*}

\begin{table}
\centering
\caption[]{The average fractional deviation,
  $\left< \left( \kappa_{\rm max} - \kappa_{\rm p} \right)/\kappa_{\rm p} \right>$,
  between the convergence of main lens plane and lensing peaks for sources at $z_{\rm s} = 1.0$.}
\label{Tab1}
\begin{tabular}{c|c|c}
\hline                                         & noise-free           & noisy                \\
\hline
%\hline $1\sigma \leq \kappa_{\rm p} < 3\sigma$ & $-0.19\ (\pm\ 0.30)$ & $-0.73\ (\pm\ 0.22)$ \\
\hline $3\sigma \leq \kappa_{\rm p} < 5\sigma$ & $-0.172\ (\pm\ 0.224)$ & $-0.678\ (\pm\ 0.242)$ \\
\hline $\kappa_{\rm p} \geq 5\sigma$           & $-0.110\ (\pm\ 0.170)$ & $-0.378\ (\pm\ 0.254)$ \\
\hline
\end{tabular}
\end{table}

Before we explore the correspondence between the lensing peaks and collapsed
massive haloes along the LOS, we first examine the contribution from different
lens planes for a given peak in the weak lensing maps. Fig.~\ref{Fig4} shows
the PDF of the main lens planes for lensing peaks with different heights in the
noise-free (left panel) and noisy (right panel) convergence fields for sources
at $z_{\rm s} = 1.0$, where the main lens plane is the plane (with convergence
$\kappa_{\rm max}$) which contributes mostly to a given lensing peak $\kappa_{\rm p}$.
The black solid and red solid lines are for peaks with
$0.054\ (3\sigma) \leq \kappa_{\rm p} < 0.089\ (5\sigma)$ and
$\kappa_{\rm p} \geq 0.089\ (5\sigma)$ respectively. Note that in the following,
we will call the peaks in these two ranges as lower and high peaks unless otherwise
specified. In table~\ref{Tab1}, we estimate the average fractional deviation
between the main lens plane and convergence peak in weak lensing maps as
$\left< \left( \kappa_{\rm max} - \kappa_{\rm p} \right) /\kappa_{\rm p} \right>$.
In the noise-free fields, we find that lensing peaks can be dominated by only
one lens plane, especially for peaks in $\kappa_{\rm p} \geq 5\sigma$, with an
average fractional deviation of $-11\%$  as show in table~\ref{Tab1}.
After adding the Gaussian shape noise, one noticeable
feature seen from the right panel is that the PDF of fractional difference
are significantly skewed. For high peaks in the range of $\kappa_{\rm p} \geq 5\sigma$,
the deviation between $\kappa_{\rm max}$ and $\kappa_{\rm p}$ becomes larger,
which implies that noise can boost peak heights in convergence maps
\citep[e.g.,][]{2010ApJ...719.1408F, 2011PhRvD..84d3529Y}.

In Fig.~\ref{Fig4} we also show the cumulative distribution functions (CDF) of the
fractional difference using the dot-dashed lines. We define a peak to be single
plane dominated peak (SPDP) if the main plane with $\kappa_{\rm max}$ larger
than $50\%$ of the total peak height $\kappa_{\rm p}$. It is found that in the noise-free field $\sim 95\%$
of high peaks are SPDPs. While this fraction decreases to $\sim 87\%$ for the
lower peaks. Adding random noise can bias the measurements of peak counts
significantly as displayed in the right panel of Fig.~\ref{Fig4}. Only $\sim 18\%$
of lower peaks are SPDPs and about $62\%$ for high peaks. Note that in this part
we do not distinguish the contribution from dark matter haloes or from the structure
around haloes, such as filaments or sheet. In the next section, we will focus on
the contribution from the dark matter haloes.

%%%%%%%%%%%%%%%%%%%%%%%%%%%%%%%%%%%%%%%%
\subsection{Correspondence between Convergence Peaks and Massive Haloes}
\begin{figure}
  \centering
  \includegraphics[width=0.475\textwidth, angle=0]{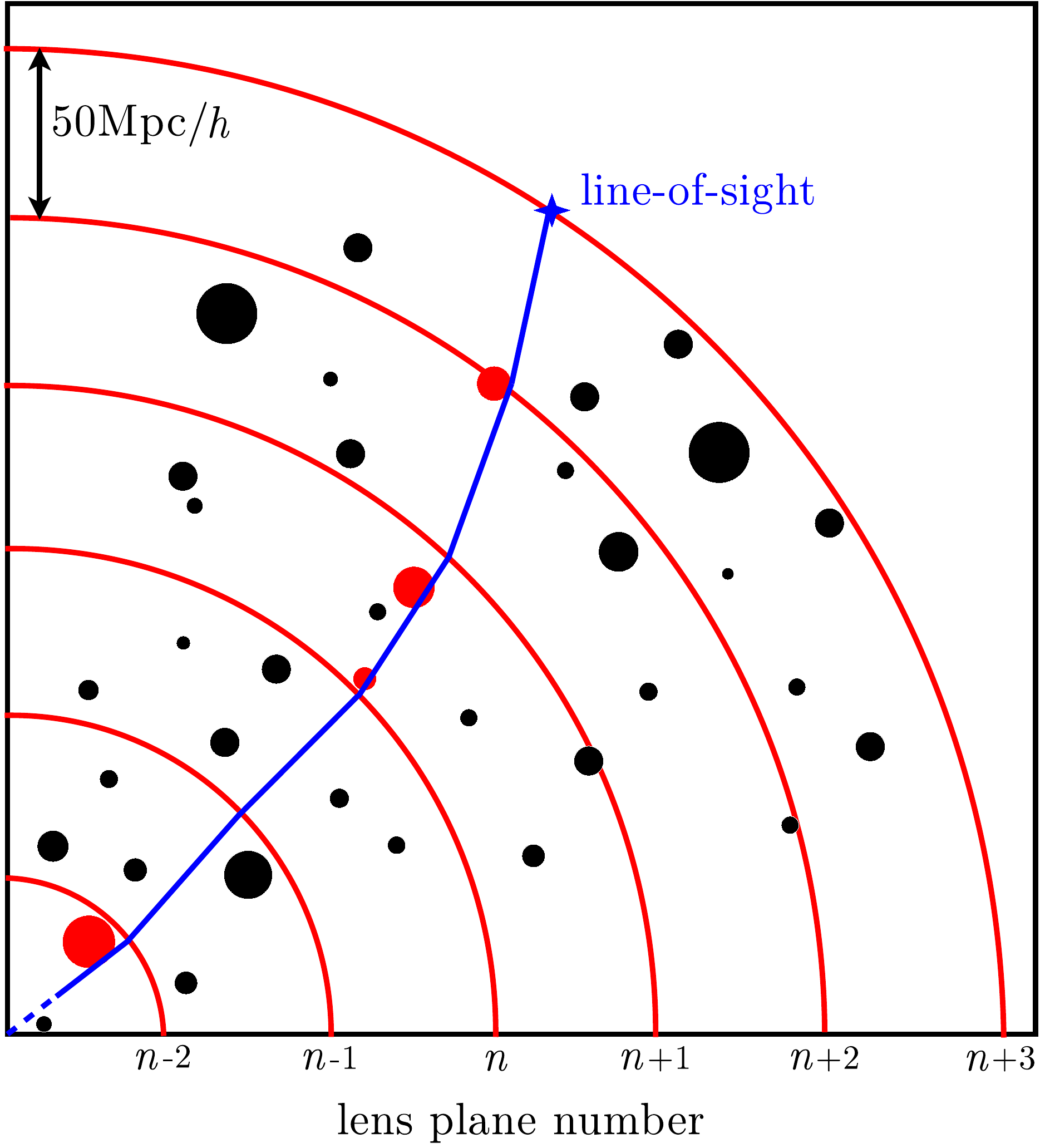}
  \caption{A cartoon picture to illustrate how we search the matched dark
  matter haloes for a given lensing peakr. Following our ray-tracing simulation, 
  light beam (blue line) can be gravitationally deflected at the position on the
  lens plane (red curves). In the picture the effect of lensing deflection is exaggerated.
  Black circles indicate the dark matter haloes
  reside in the light-cone. Red circles show the matched haloes if the traced
  light of the convergence peak crosses the virial radius of these haloes.
}
  \label{Fig5}
\end{figure}

\begin{figure*}
  \centering
  \includegraphics[width=0.85\textwidth, angle=0]{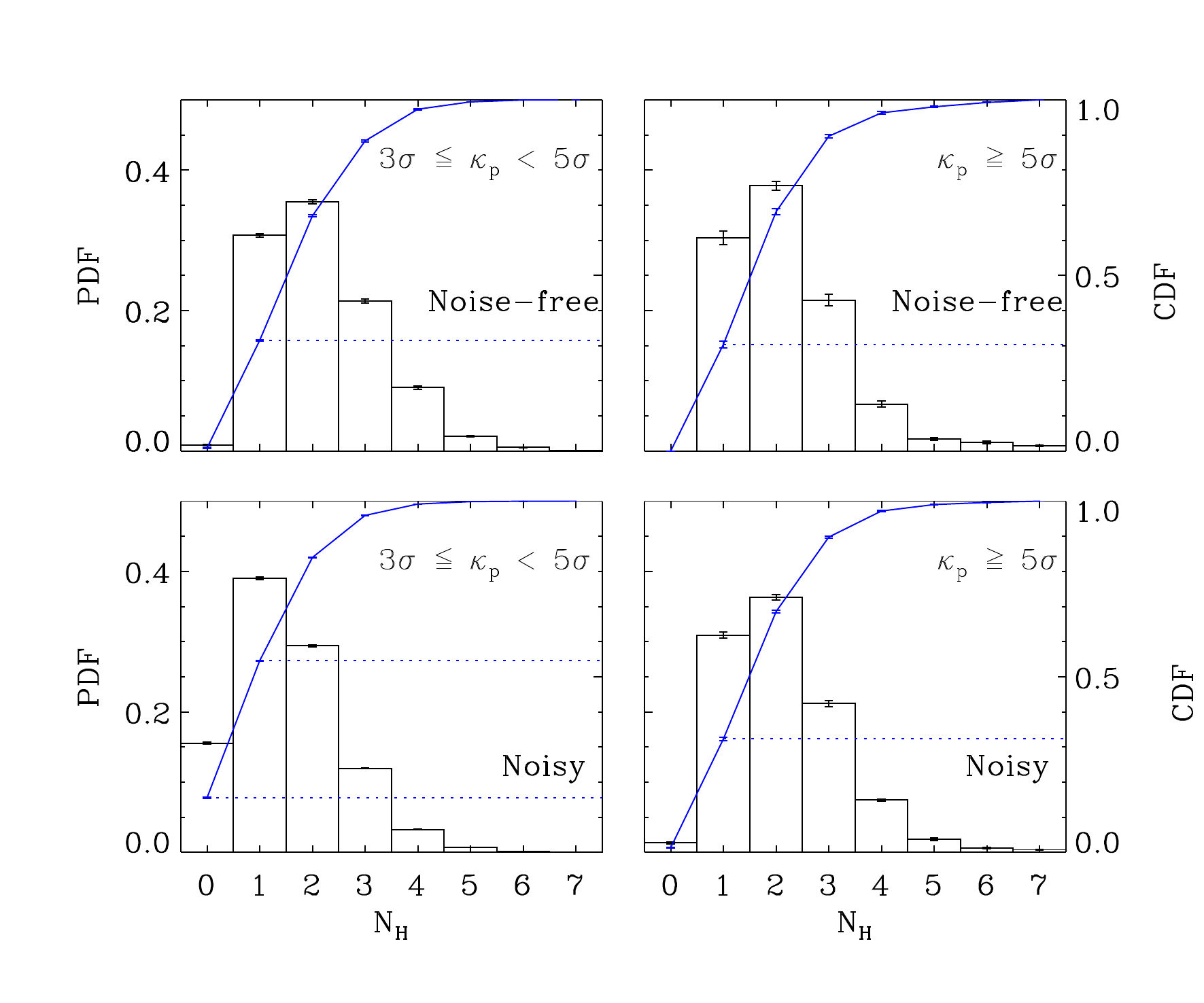}
  \caption{The  number  of  dark matter haloes contributing to a given
  convergence  peak in  the  noise-free fields (upper two panels) and
  noisy  fields (lower two panels). Here we consider two samples with
  different peak heights, $3\sigma \leq \kappa_{\rm p} < 5\sigma$ and
  $\kappa_{\rm p} \geq 5\sigma$, respectively. In addition, blue lines
  show the CDF of the number of the related haloes.}
  \label{Fig6}
\end{figure*}

In general, convergence peaks are associated with collapsed haloes along
the LOS, and it is also clear that this correspondence will be affected
by the projection of LSS, noise, and also the observational effects, such
as masking effects \citep[e.g.,][]{2002ApJ...575..640W, 2010ApJ...719.1408F, 2014ApJ...784...31L,
2015MNRAS.453.3043S, 2018ApJ...857..112Y}. The height of convergence peak is closely
related to the mass of the dark matter haloes along the LOS \citep[e.g.,][]{
2004MNRAS.350..893H}. In this section and Sec. 4.3, we focus on the
contribution from massive haloes with $M_{\rm H} \geq 10^{13}{\rm M}_{\sun}$,
which can produce a peak with a significance greater than $1\sigma$ in the
convergence maps for sources at $z_{\rm s} = 1.0$. In Sec. 4.4, we will
focus on more massive haloes. Fig.~\ref{Fig5} shows a cartoon picture to
illustrate how we search the matched dark matter haloes for a given lensing
peak. In the picture, red curves represent the configuration of spherical
lens planes. Following our ray-tracing simulation, light beam (blue line)
is gravitationally deflected at the position on the lens plane. Black and
red circles indicate the dark matter haloes reside in the light-cone. For
each lensing peak, we then find all matched dark matter haloes if the traced
light of the convergence peak crosses the virial radii of these haloes (red circles).

In Fig.~\ref{Fig6} we show the PDF of the number of haloes which are related
to a given peak. Error bars are estimated by the Jackknife method. In the upper
two panels, peaks are extracted from the noise-free convergence fields and lower
panels show the results in the noisy convergence fields. For lensing peaks in
the noise-free maps, we find that a large fraction ($> 65\%$) of peaks can be
related to more than one dark matter haloes  along  the line of sight. This
result is consistent with previous work. In \citet{2011PhRvD..84d3529Y}, they
examine the number of haloes required to account for $> 50\%$ of the peak height,
and found $\sim 55\%$ high peaks for sources at $z_{\rm s} = 2$ with $\kappa \geq 0.11$
are contributed by more than one massive haloes on the line of sight. Moreover,
as shown in the upper-left panel of Fig.~\ref{Fig6}, a few ($\lesssim 1\%$) of
lower peaks have no matched massive haloes along the sightline. This can
be explained by the projection of LSS or by the low mass haloes
($M_{\rm H} < 10^{13}{\rm M}_{\sun}$) along the LOS in our simulation. Note that
in this work we only focus on massive haloes ($\geq 10^{13}{\rm M}_{\sun}$) and
we do not distinguish the contribution on convergence from low mass haloes
($< 10^{13}{\rm M}_{\sun}$) with that from diffuse dark matter particles, or
those from the LSS, such as filament or sheets. For simplicity, we use the LSS
term to denote all these structures other than the massive haloes. After adding
noise, more lensing peaks have no matched haloes on the line of sight in both
peak height range. For high peaks, we find $\sim 2\%$ of lensing peaks without
matched massive haloes. Compared with the upper-right panel, we conclude that
these high peaks arise from the boosting effects of the noise field.
While for lower peaks the PDF distributions imply that $\sim 15\%$ of false peaks
is caused by the noise, and the fraction of peaks which relate to more than one
dark matter haloes decrease to $\sim 45\%$.

\begin{figure*}
  \centering
  \includegraphics[width=0.95\textwidth, angle=0]{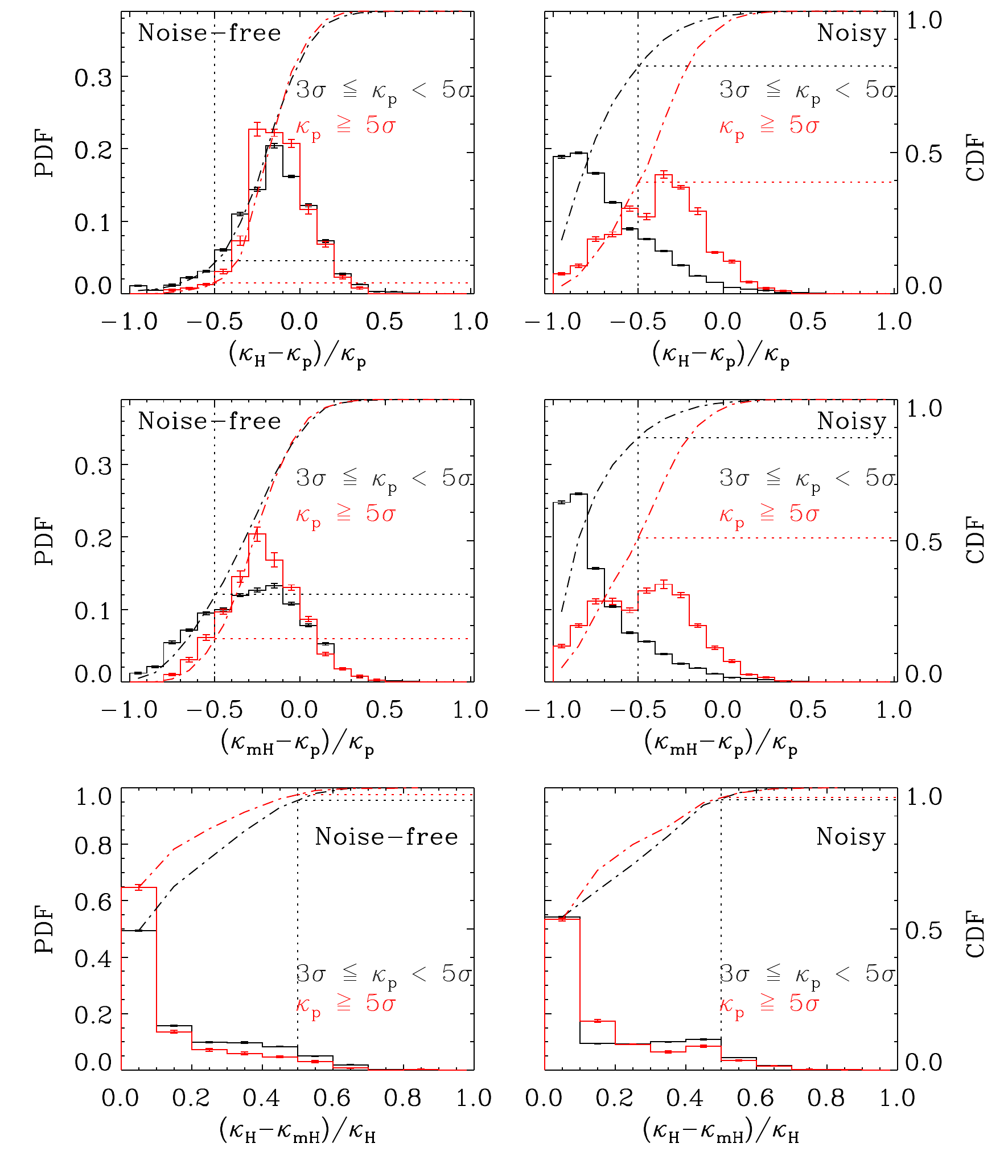}
  \caption{The top panels show the probability distribution functions of the difference
  between halo contribution $\kappa_{\rm H}$ and lensing peak $\kappa_{\rm p}$ in the
  noise-free (left panel) and noisy (right panel) convergence maps. The fractional difference
  are calculated by $\left( \kappa_{\rm H} - \kappa_{\rm p} \right)/\kappa_{\rm p}$ for
  two different samples, $3\sigma \leq \kappa_{\rm p} < 5\sigma$ and $\kappa_{\rm p} \geq 5\sigma$,
  respectively. The middle panels show the fractional difference between the lensing peak and the
  prediction of the only main halo along the line of sight, 
  $\left( \kappa_{\rm mH} - \kappa_{\rm p} \right)/\kappa_{\rm p}$. 
  The bottom panels show the fractional differences between $\kappa_{\rm mH}$ and $\kappa_{\rm H}$ 
  for the given peaks.
  In addition, dot-dashed lines show the CDF of the fractional difference in each case.}
  \label{Fig7}
\end{figure*}

\begin{table}
\centering
\caption[]{The average fractional deviation,
  $\left< \left( \kappa_{\rm H} - \kappa_{\rm p} \right)/\kappa_{\rm p} \right>$,
  between the halo prediction and lensing peak for sources at $z_{\rm s} = 1.0$.}
\label{Tab2}
\begin{tabular}{c|c|c}
\hline
                                               & noise-free           & noisy                \\
\hline
\hline $3\sigma \leq \kappa_{\rm p} < 5\sigma$ & $-0.169\ (\pm\ 0.250)$ & $-0.718\ (\pm\ 0.268)$ \\
\hline $\kappa_{\rm p} \geq 5\sigma$           & $-0.128\ (\pm\ 0.176)$ & $-0.394\ (\pm\ 0.271)$ \\
\hline
\end{tabular}
\end{table}

\begin{figure*}
  \centering
  \includegraphics[width=0.98\textwidth, angle=0]{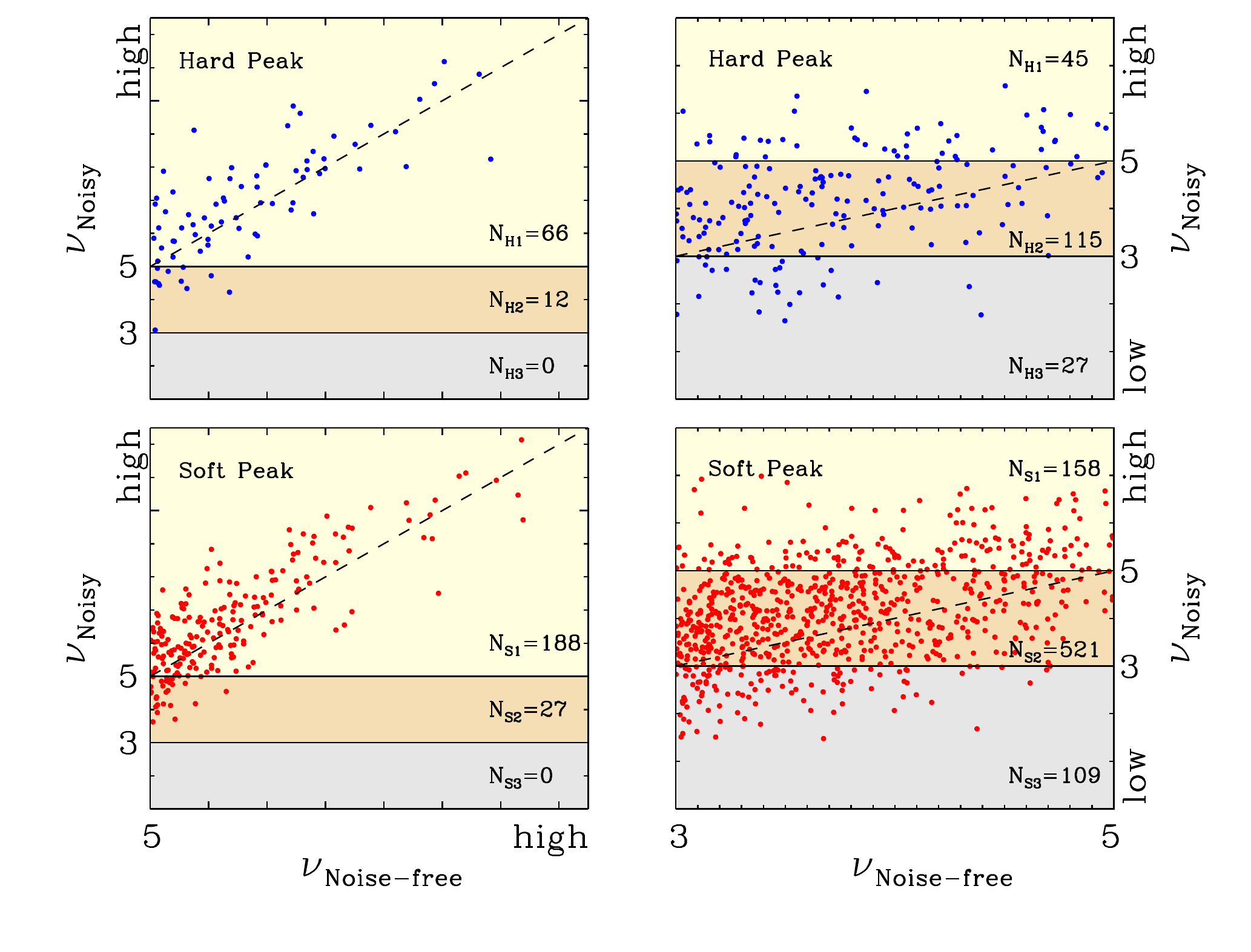}
  \caption{The distribution of the peak heights for the true peaks.
           Blue circles denote hard peaks and red circles are soft
           peaks. The numbers of peaks in different ranges of SNR
           are given by $N_{{\rm H}i}$ and $N_{{\rm S}i}$ respectively.}
  \label{Fig8}
\end{figure*}

In order to study the contribution of massive haloes to a given lensing peak,
we use Eq.~\ref{equ:kNFW} to predict the value of a peak height, $\kappa_{\rm H}$, 
by assuming that FOF haloes are described by the truncated NFW density profiles.
In the top panels of Fig.~\ref{Fig7}, we show the difference
between halo contribution $\kappa_{\rm H}$ and the total lensing peak $\kappa_{\rm p}$
in the noise-free (upper-left panel) and noisy (upper-right panel) convergence
maps. It is seen that in case of no noise, the contribution is similar for high
and lower peaks. In table~\ref{Tab2} we give the average fractional deviation
between halo predictions and convergence peaks,
$\left< \left( \kappa_{\rm H} - \kappa_{\rm p} \right)/\kappa_{\rm p} \right>$.
For high peaks in the noise-free field, an average fractional deviation of $-0.128$
is obtained, which means that LSS contributes 12.8 per cent to the lensing peaks
with $\kappa_{\rm p} \geq 0.089$ for sources at $z_{\rm s} = 1.0$. The contribution
of LSS is increased to $16.9\%$ for lower peaks.

The right top panel show the case of noisy convergence field. Compared to the
left top panel, it is seen that the distributions of both high and lower peaks
are skewed to the left, with more skewness for lower peaks, indicating a significant
fraction of peaks are false, and they are not related to any massive haloes.
Table~\ref{Tab2} shows that with noise the average fractional deviation is $-0.394$
for high peaks, and is $-0.718$ for lower peaks. The CDF distributions (dot-dashed
lines) show that in the noise-free case, $96\%$ of high peaks are dominated by
massive haloes which contribute more than half of the total peak height, and this
fraction decreases to $88.4\%$ for lower peaks. In the case of noise, this fraction
decreases to $60\%$ and $17\%$ for high and lower peaks.

\begin{table}
\centering
\caption[]{The average fractional deviation,
  $\left< \left( \kappa_{\rm mH} - \kappa_{\rm p} \right)/\kappa_{\rm p} \right>$,
  between the prediction of the main halo and lensing peak for sources at $z_{\rm s} = 1.0$.}
\label{Tab3}
\begin{tabular}{c|c|c}
\hline
                                               & noise-free           & noisy                \\
\hline
\hline $3\sigma \leq \kappa_{\rm p} < 5\sigma$ & $-0.297\ (\pm\ 0.290)$ & $-0.767\ (\pm\ 0.241)$ \\
\hline $\kappa_{\rm p} \geq 5\sigma$           & $-0.219\ (\pm\ 0.215)$ & $-0.472\ (\pm\ 0.281)$ \\
\hline
\end{tabular}
\end{table}

Furthermore, we examine whether these lensing peaks are dominated by
only one massive halo along the line of sight. In the middle panels
of Fig.~\ref{Fig7} we show the distribution of fractional difference
between the convergence peaks and the main dark matter haloes,
$\left( \kappa_{\rm mH} - \kappa_{\rm p} \right)/\kappa_{\rm p}$.
Here the main dark matter halo is the halo contributing mostly to the
given peak. The distributions are similar to those in the upper panels.
Table~\ref{Tab3} shows that the average fractional deviation in the
noise-free field is to $21.9\%$ and $29.7\%$ for high and lower peaks
respectively. With noise, the average fractional deviation is much worse.
Similar as the definition of SPDP, we define a peak is a one halo
dominated peak (OHDP) if $\kappa_{\rm mH} \geq 0.5 \kappa_{\rm p}$.
It is found that without noise, $84.7\%$ of high peaks are OHDPs and
$69\%$ for lower peaks, indicating that without noise most high peaks
are dominated by single massive halo, consistent with previous work
\citep[e.g.,][]{2004MNRAS.350..893H, 2011PhRvD..84d3529Y}. However,
noise will significantly contaminate the peaks. For example only
$\sim 48\%$ and $\sim 12\%$ of high and lower peaks are dominated by
one single massive halo in case of noise. 
Moreover, the bottom panels show the fractional differences
between $\kappa_{\rm mH}$ and $\kappa_{\rm H}$ for the given peaks.
It is seen that the halo predictions are mostly contributed by the 
main haloes for both the noise-free and noisy cases.
Here we note that these number are only for our selection of dark 
matter halo mass and our modeling of noise with its height and distribution.

\begin{figure*}
  \centering
  \includegraphics[width=0.98\textwidth, angle=0]{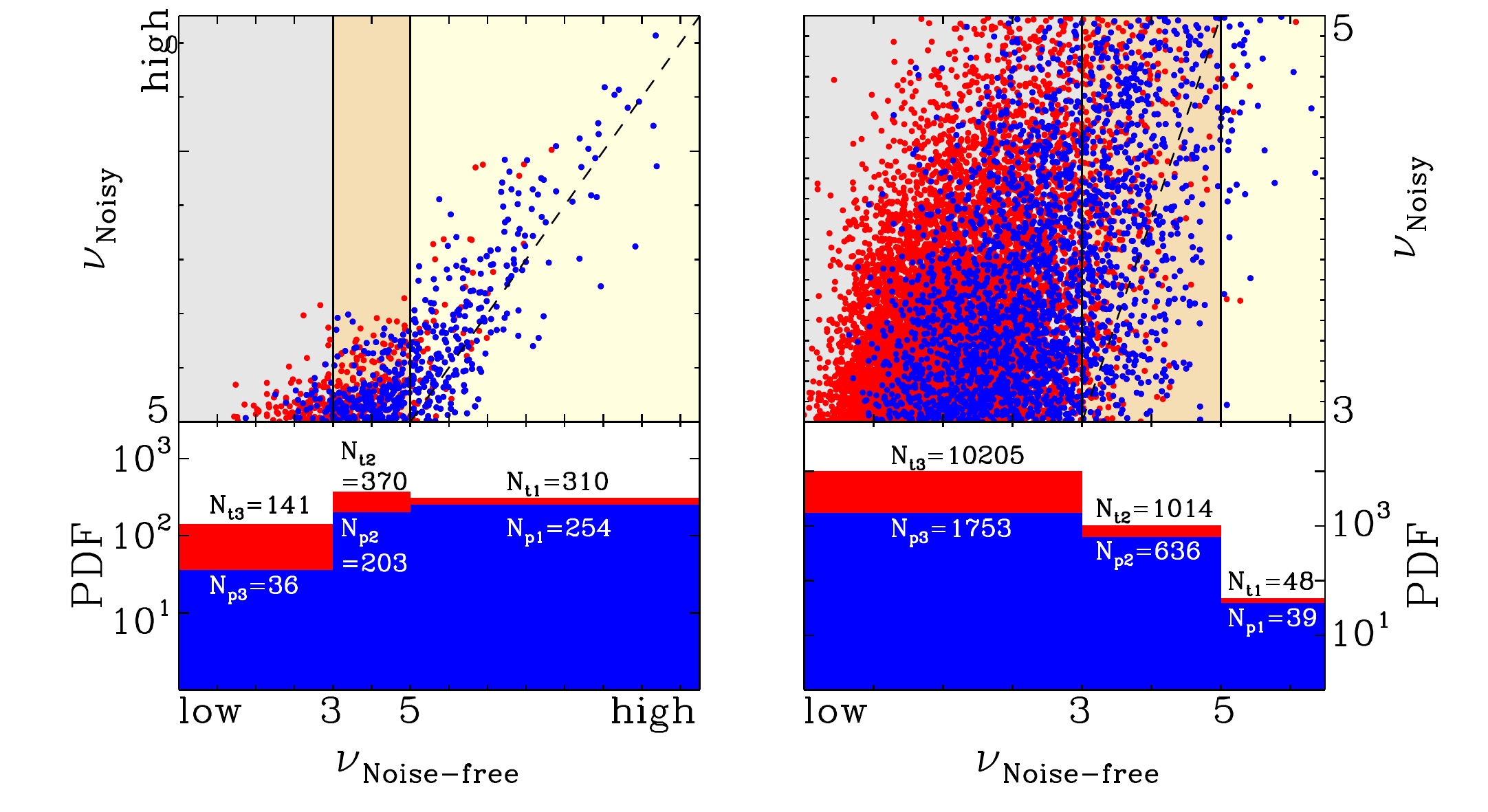}
  \caption{The SNR distribution of convergences peak in the noisy
           field. Red circles denote all identified peaks in the
           noisy field, and blue circles are the matched true peaks.
           In the bottom panels, the PDFs show the number of peaks
           in the different range of SNR. $N_{{\rm t}i}$ denotes the
           total number of peaks in the noisy field, and $N_{{\rm p}i}$
           is the number of matched peaks from the noise-free field.}
  \label{Fig9}
\end{figure*}

%%%%%%%%%%%%%%%%%%%%%%%%%%%%%%%%%%%%%%%%
\subsection{Effects of Noise on the Convergence Peak and Halo Connection}
\begin{table}
\centering
\caption[]{Counts of peaks in our convergence field.
           The numbers in parentheses in the middle
           and right columns show the number of peaks
           per 1 ${\rm deg}^{2}$ in our mocked sky.}
\label{Tab4}
\begin{tabular}{clll}
  \hline
	    & \multirow{2}{*}{Case} & \multicolumn{2}{c}{Number} \\
        &                       & high peaks & lower peaks \\
  \hline
    I   & peak in noise-free field &  407 (0.21) &  2095 (1.07) \\
    II  & peak in noisy field  & 821 (0.42) & 11267 (5.75) \\
	%iii & peak in both field & 78 (0.04)   &  187 (0.10)  \\
	%iv  & peak in noise-free field, but not in noisy field &  329 (0.17)  &  1908 (0.97)   \\
	%v   & peak in noisy field, but not in noise-free field &  706 (0.36)  &  10911 (5.57)   \\
    III  & peak in the pure noise field  &  2 (0.001)  &  3898 (1.99) \\
  \hline
\end{tabular}
\end{table}

The above peak statistics analyses show that convergence peaks
can be significantly affected by galaxy shape noise. In \citet{2010ApJ...719.1408F},
they developed a theoretical model to predict the numbers of
convergence peaks taking into account the effects of noise
in the weak lensing maps. They concluded that the noise not
only produces pure noise peaks in the convergence field, but also
affects the heights of true peaks. In addition, the pure noise peaks
will be enhanced in halo regions (the virial radius). In this work
we further study the effects of noise on the correspondence between
the peaks and massive haloes. Table~\ref{Tab4} shows the peak counts
in our convergence field for the high and lower peaks. It is found
that without noise, the number density of high and lower peaks is
0.21 ${\rm deg}^{-2}$ and 1.07 ${\rm deg}^{-2}$ respectively.
Noise will double the number of high peak to 0.42 ${\rm deg}^{-2}$,
and raise the number of lower peaks to 5.75 ${\rm deg}^{-2}$.
The number density of peaks in the noisy field is consistent with
that of \citet{2018arXiv180101886G}.

In order to examine the effect of noise on the true peaks,
we then check how many true peaks survived and how much of
their SNR change. Here we distinguished two cases. For a peak
in the noise-free convergence field, if it is still a peak in
the noisy field, we call it as a hard peak. If the true peak
disappear in the noisy field, but its neighbor (the angular
distance is less than one $\rm arcmin$, more than twice of
the resolution of our simulation with 0.43 $\rm arcmin$) appears
as a peak, we call the peak in the noise-free field as a soft peak.
We call these two kinds of peak as true peak and the  others
noise-free peaks are defined as destroyed peaks.
Note that true peaks are defined as the local maxima in the noise-free 
convergence field. Thus a fraction of true peaks are caused by projection 
effects. This projection effects of LSS have been shown in 
Fig.~\ref{Fig6} and Fig.~\ref{Fig7}.

Fig.~\ref{Fig8} shows the comparison of SNR of the peak pairs.
The upper panels are for hard peaks and lower for soft peaks.
The numbers of hard/soft peaks in different ranges of SNR are 
given by $N_{{\rm H}i}$ and $N_{{\rm S}i}$ respectively.
In the $\sim 2000\ {\rm deg}^2$ of our mock sky, we find a total
number of 407 and 2095 for the high and lower peaks in the noise-free
field (see table~\ref{Tab4} for those numbers). For those high peaks,
it is found that the total number of survived peaks is
$N_{\rm H1}+N_{\rm H2}+N_{\rm H3}+N_{\rm S1}+N_{\rm S2}+N_{\rm S3} = 293$,
about $72\%$ of the original high peaks. The remaining $114$ peaks, $28\%$
of the total, are destroyed by noise. Among those survived peaks,
$78\ (27\%)$ peaks are hard peaks and others are soft peaks. Among
those hard high peaks, about $85\%$ are still high peaks, but others
are downgraded to low peaks. In the right panel we show the case of
the low peaks. It is seen that only $47\%$ ($975$ of $2095$) peaks
survive the adding of noise. More than half of original peak are
destroyed.

Previously we search the noise-free field and find their correspondences
in the noisy field. Now we start from the noisy field and find their
correspondences in the noise-free field. Fig.~\ref{Fig9} shows the SNR distribution
of peaks identified in the noisy field, and the PDF in the bottom panels
show the number of peaks in the given ranges of SNR, where $N_{{\rm t}i}$
denotes the total number of convergence peak in the noisy field and $N_{{\rm p}i}$
is the number of matched peaks (with $N_{\rm p1}$/$N_{\rm p2}$ denoting the original
high/lower peaks) from the noise-free field. Thus the matched
peaks here contain both hard peaks and soft peaks in Fig.~\ref{Fig8}, which
means $N_{\rm p1} = N_{\rm H1}+N_{\rm S1}$ for high peaks and $N_{\rm p2} = N_{\rm H2}+N_{\rm S2}$
for lower peaks. For the $821$ high peaks in the noisy field,
$493\ (60\%)$ of them are true peaks in the noise-free field, and among which
$254$ (around $254/493 \sim 51\%$) are original high peaks. The remaining
$40\%$ are not peaks in the noise-free field, but are boosted as peaks by
noise. In the right panel, it is found that there is $11267$ lower peaks
in the noisy field, but only $20\%\ (2428/11267)$ are true peaks in the
noise-free convergence field. These results suggest that for high peaks
in the noisy field there are significant fraction of them are true peaks
in the density map, and around $30\%$ are true high peaks.

%%%%%%%%%%%%%%%%%%%%%%%%%%%%%%%%%%%%%%%%
\subsection{Mass Dependence of the Correspondence between Peaks and Haloes}
\begin{figure*}
  \centering
  \includegraphics[width=0.95\textwidth, angle=0]{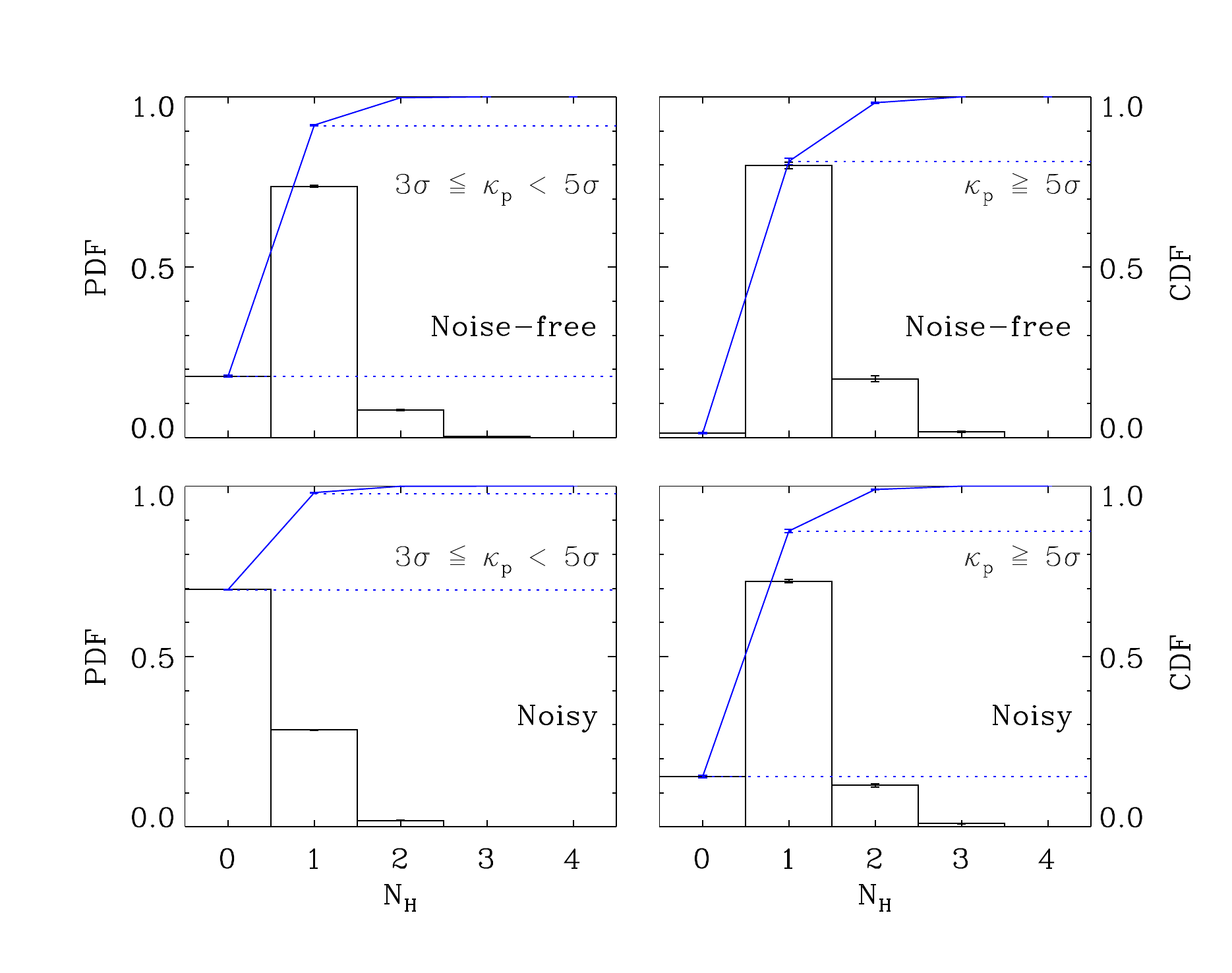}
  \caption{Similar as Fig.~\ref{Fig6}, but only with halo mass
           $M_{\rm H} \geq 10^{14} {\rm M}_{\sun}$.}
  \label{Fig10}
\end{figure*}

As noted in Sec. 4.1 we only consider haloes with mass larger than
$10^{13} {\rm M}_{\sun}$ and Fig.~\ref{Fig7} shows that most of high
peaks are dominated by a main halo along the LOS (OHDPs).
In this section, we only take  account  of  dark  matter
haloes  with $M_{\rm H} \geq 10^{14} {\rm M}_{\sun}$ to examine the
correspondence between convergence peaks and dark matter haloes.

\begin{figure*}
  \centering
  \includegraphics[width=0.95\textwidth, angle=0]{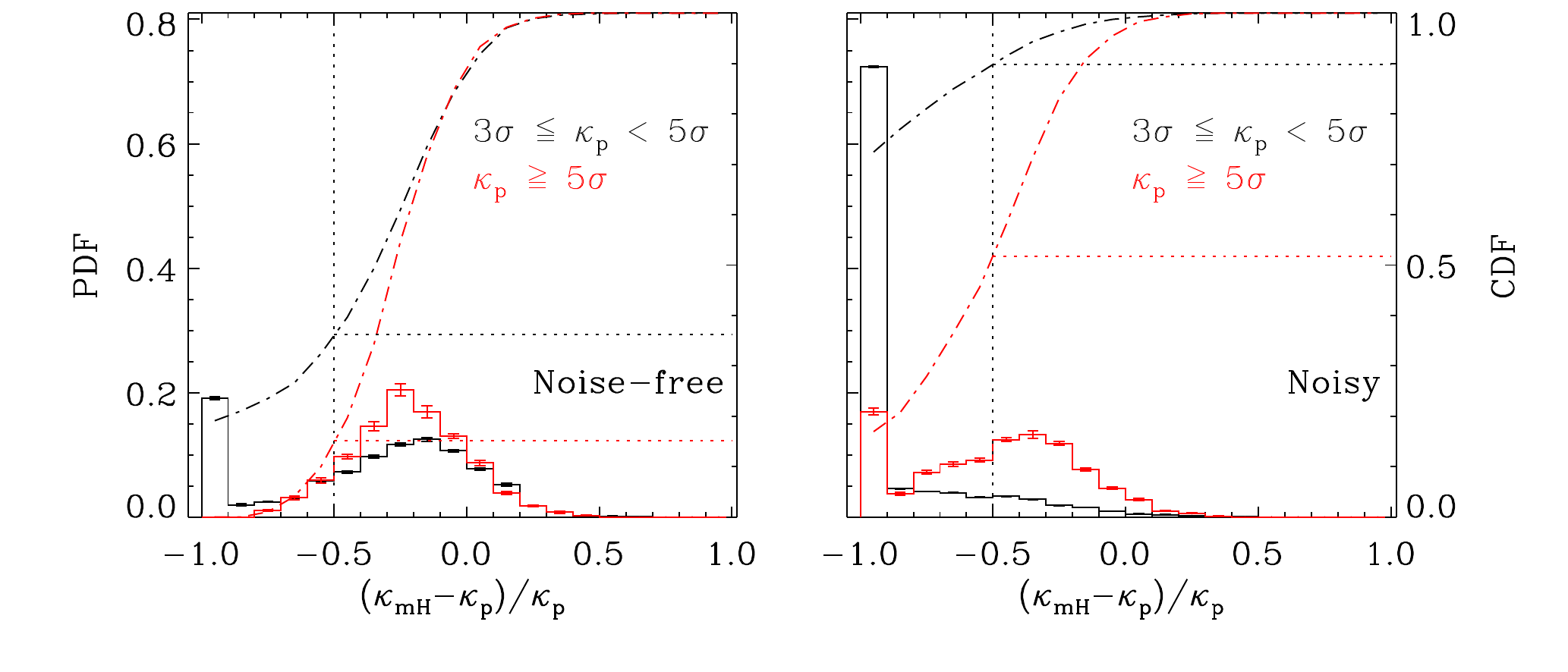}
  \caption{Similar as Fig.~\ref{Fig7}, but only with halo mass
           $M_{\rm H} \geq 10^{14} {\rm M}_{\sun}$.}
  \label{Fig11}
\end{figure*}

Similar as Fig.~\ref{Fig6} and Fig.~\ref{Fig7}, the results are
given in Fig.~\ref{Fig10} and Fig.~\ref{Fig11}. As shown in
Fig.~\ref{Fig10}, we display the related halo numbers in the
noise-free (upper) and noisy field (lower). Comparing the upper-right
panels of Fig.~\ref{Fig6} and Fig.~\ref{Fig10}, we conclude most
of high peaks in noise-free field can be related with only one massive
halo larger than $10^{14} {\rm M}_{\sun}$, and about $\sim 20\%$
of them are associated with more than one massive haloes larger than
$10^{14} {\rm M}_{\sun}$ along the LOS. For the lower peaks in the
upper-left panel of Fig.~\ref{Fig10}, we find that $18\%$ of peaks
do not have matched haloes with mass larger than $10^{14} {\rm M}_{\sun}$.
Compared with Fig.~\ref{Fig6}, these not matched peaks are attributed to
one or more haloes with mass in $10^{13} {\rm M}_{\sun} \leq M_{\rm H} < 10^{14} {\rm M}_{\sun}$.
After adding noise, about $70\%$ high peaks can relate to only one
massive halo (lower-right), but $70\%$ of lower peaks do not have
matched massive haloes (lower-left), which can be related with less
massive dark matter haloes or caused by the noise ($\sim 15\%$). By
comparing with the results of Fig.~\ref{Fig6}, we find that massive
haloes with $M_{\rm H} \geq 10^{14} {\rm M}_{\sun}$ can be effectively
detected by high peaks with $\kappa_{\rm p} \geq 5\sigma$, but only
about $30\%$ lower peaks are related with such massive haloes.

Furthermore, Fig.~\ref{Fig11} shows the fractional difference
between convergence peaks and the main halo as in Fig.~\ref{Fig7}.
In the noise-free field (left panel), the average fractional
deviation is $-0.218$ for the high peaks, which is almost same
as that in table~\ref{Tab3}. This consistency means that high
peaks are dominated by massive haloes with $M_{\rm H} \geq 10^{14} {\rm M}_{\sun}$.
For lower peaks, the deviation increase to $-0.376$, which is
slightly larger than that in table~\ref{Tab3}. This is reasonable
as that we find $\sim 18\%$ of lower peaks do not have matched
haloes with $M_{\rm H} \geq 10^{14} {\rm M}_{\sun}$ in Fig.~\ref{Fig10}.
These peaks should be explained by the contribution of lower
mass haloes in $10^{13} {\rm M}_{\sun} \leq M_{\rm H} < 10^{14} {\rm M}_{\sun}$.

The distribution of CDF in Fig.~\ref{Fig11} shows almost same
results as in Fig.~\ref{Fig7}, which means dark matter haloes
with $M_{\rm H} \geq 10^{14} {\rm M}_{\sun}$ should be the
main contributor to the convergence peaks although lower mass
haloes are also important to give $\sim 5\%$ difference for
lower peaks. After adding noise, we find about 50\% high peaks
can be dominated by the massive haloes with $M_{\rm H} \geq 10^{14} {\rm M}_{\sun}$,
but only 10\% lower peaks are dominated by one massive halo
along the LOS. This effective mass detection in
weak lensing convergence peaks is also found in the
literature \citep[e.g.,][]{2004MNRAS.350..893H, 2018MNRAS.474.1116S}.

%%%%%%%%%%%%%%%%%%%%%%%%%%%%%%%%%%%%%%%%
\section{CONCLUSION AND DISCUSSION}
In this paper, we focus on lensing peaks to examine the correspondence between weak lensing convergence peaks
and dark matter haloes. Following the multiple lens plane algorithm on the curved sky, We performed a high-spatial
and high-mass resolution ray-tracing simulation. Based on these simulated convergence maps, we examine the abundance
of lensing peaks and the correspondence between peaks and massive haloes along the LOS. For each lensing peak, we
try to search the corresponding dark matter haloes, if the selected peak locates in the virial radius of these haloes.
Two thresholds, $\kappa = 3\sigma(0.054)$ and $5\sigma(0.089)$, are used in our analysis and the results are
summarized as follows.

\begin{itemize}
\item[(1)] In the noise-free field, we find that more than $65\%$ peaks are related to more than one massive
haloes with mass in $M_{\rm H} \geq 10^{13}{\rm M}_{\sun}$. A few ($\lesssim 1\%$) of lower peaks in
$3\sigma \leq \kappa_{\rm p} < 5\sigma$ arise without matched massive haloes along the LOS. This should be
contributed by the projection effect of LSS. By assuming these matched haloes with the truncated NFW profile,
we then test the correspondence of peak heights between weak lensing maps and the related massive haloes. The
fractional difference between halo predictions and lensing peaks, $(\kappa_{\rm H} - \kappa_{\rm p})/\kappa_{\rm p}$,
shows that LSS contributes $12.8\%$ to the lensing peaks with $\kappa_{\rm p} \geq 0.089$ for sources at $z_{\rm s} = 1.0$.
The contribution from LSS increase to $16.9\%$ for lower peaks. On other hand, as shown by the cumulative
distribution functions, about $96\%$ peaks is dominated by the massive dark matter haloes along the LOS for
high peaks, and for lower peaks the fraction is about $88.4\%$. Moreover, we further examine the correspondence
between the lensing peaks and the main dark matter haloes. The result shows that $84.7\%$ high peaks can be
dominated by one single halo and only $69\%$ for the lower convergence peaks.

\item[(2)] In the noisy field, convergence maps are significantly modulated by the noise and the asymmetric
double peaks arise due to the noise modulation. After adding noise, a number of false peaks arise, especially
at the lower SNR. About $60\%$ high peaks are true peaks, but only $20\%$ for the lower peaks. The average
fractional deviation show that for high peaks the difference between convergence peaks and the predictions
from total matched haloes is $39.4\%$ and $60\%$ of them are halo dominated, but for lower peaks only $17\%$ are
halo dominated. If we only consider the main dark matter halo with the maximum contribution in the halo
prediction, we find the average fractional deviation slightly increase to $47.2\%$ and the fraction of peaks
dominated by the main halo decrease to $48\%$.
\end{itemize}

Furthermore, considering that most of high peaks are OHDPs, we take account of more massive haloes with
$M_{\rm H} \geq 10^{14} {\rm M}_{\sun}$ to examine the correspondence between convergence peaks and dark
matter haloes. We find dark matter haloes with $M_{\rm H} \geq 10^{14} {\rm M}_{\sun}$ are the main contributor
to the convergence peaks although lower mass haloes are also important to give $\sim 5\%$ difference for
lower peaks. After adding noise, we find about $50\%$ high peaks are dominated by the massive haloes with
$M_{\rm H} \geq 10^{14} {\rm M}_{\sun}$, but only $10\%$ lower peaks are dominated by one massive halo
along the LOS. Our results show that massive haloes, especially with $M_{\rm H} \geq 10^{14} {\rm M}_{\sun}$,
can be effectively detected by the high weak lensing convergence peaks. In the lower peaks with
$3\sigma \leq \kappa < 5\sigma$, the peak heights are significantly modified by the noise effect. When we use
it to probe cosmology and constrain the model of structure formation, we should carefully address the effects
of galaxy shape noise, LSS projection and also the observational effects, especially for low peaks.

%%%%%%%%%%%%%%%%% Acknowledgement
\section*{Acknowledgements}
This work is supported by the NSFC (No.11333008, 11273061, 11333001, 11653001),
the 973 program (No. 2015CB857003, 2013CB834900), and the NSF of Jiangsu province
(No. BK20140050). XKL acknowledges the support from YNU Grant KC1710708.

%\bibliographystyle{mnras}
%\bibliography{ref} % if your bibtex file is called example.bib


%%%%%%%%%%%%%%%%%%%%%%%%%%%%%%%%%%%%%%%%%%%%%%%%%%
% Don't change these lines
\bsp	% typesetting comment
\label{lastpage}
\end{document}